# Microscopic Hamiltonian and Chaotic Behavior for Barium Titanate Nanoparticles Revealed by Photonic Tunneling Model

Emile Godwe[1], Olivia Florea[2], Rabab Jarrar[3], Mibaile Justin[4], and and Jihad Asad[3,*]

[1]Department of Physics, Faculty of Science, University of Maroua, P.O. Box 814, Cameroon

e-mail : willgod7@yahoo.com

[2]Faculty of Mathematics and Computer Science, Transilvania University of Brasov, Romania

e-mail : olivia.florea@unitbv.ro

[2]Department of Physics, Faculty of Applied Sciences, Palestine Technical University-Kadoorie, Tulkarm P 305, Palestine

e-mail: r.jarrar@ptuk.edu.ps, j.asad@ptuk.edu.ps

[4]Higher Teachers' Training College of Maroua, The University of Maroua, P.O. Box 46, Cameroon.

e-mail: thejust7@yahoo.fr

**Abstract**

In order to understand the microscopic properties of the ferroelectric nanoparticles, we review concepts of the attractive $BaTiO_3$ nanocrystals where a photon field from its constituent electrons and ions are considered. We present an associated Hamiltonian and extended potential which are the bases of the physical description of the system from certain geometry of the semiquantal space with the specific phonon mode. Furthermore, we formulate the nonlinear dynamics equation of the systems with the famous microscopic Hamiltonian and Heisenberg equation of motion. In addition, we interest ourselves to a system of $BaTiO_3$ nanocrystals and hypothesis leading to order-disorder nature of phase transitions. The latter are linked to the cooperative phenomenon of chains. The numerical simulations reveal that complex behaviors propagate in such system.

**Key words:** Semiquantal; Order-Disorder; Barium Titanate Nanocrystals; Photon Tunneling.

* Corresponding Author e-mail: j.asad@ptuk.edu.ps

# 1. Introduction

Ferroelectricity is the ability that possess certain class of materials by having an impulsive electric polarization that can reverse under the effect of an applied outward electric field [1, 2]. J. Valasek during the investigations of the dielectric properties of Rochelle salt discovered in 1921 this phenomenon [3]. It is difficult to say exactly in a single sentence what a ferroelectric is. Usually, they are viewed as materials that sustain a dielectric polarization when exposed to an electric field. Ferroelectrics are widely used because of their properties. These applications include hysteresis which is used in nonvolatile memories; a high permittivity used for capacitors engineering. These phenomena are also found in devices with resonant waves of radio frequency filters, sensors, or actuators, presenting high piezoelectric effects. Other properties are their high pyroelectric coefficients commonly used for the development of infra-red detectors [4]. Their electro-optical effects are found in optical switches, and another important characteristic is given by the abnormal resistivity coefficients. [4].

In recent years, research based on ferroelectric nanoparticles have been approached both in the field of materials technology as well as in medicine or biology. For material technology, ferroelectrics has been used in multilayered capacitors and nanocomposites [5] and in liquid crystals for display and non-display applications [6]. In biology they are used as proliferation agents for cells and tissues cultures [7, 8]. Screening for genetic diseases or cancer has an important role in the early identification of infected tissues, thus the use of ferroelectric nanoparticles in the field of medical imaging of deep tissue in vitro and in vivo represents a real success. [9] .

In ordinary materials, the polarization field is proportional to the applied electric field whereas in ferroelectric, it is a nonlinear function of the electric field then, making the ferroelectrics been nonlinear materials. Therefore, they are subject to many nonlinear phenomena. The physical phenomenon in which the physical behavior of a system cannot be correlated with the effect it generates is represented by nonlinearity. The response as well as the stimulus of such a system represent the physical parameters that are not proportional. This nonlinearity is obviously found in natural systems, their behavior being independent of their size (atom vs universe) or their type of motion (classical vs relativistic) [10]. In recent years, scientific studies have focused on chaos theory. The etymology of the term comes from the Greek language, which implies an unformed, disordered world. This expression can resemble a gaping chasm (a precipice).

[11]. From a scientific point of view, the non-linear state of a natural system is given by its non-periodic oscillatory characteristic. [12].

In this paper, we derive the extended Hamiltonian from the photonic tunneling model which Je Huan Koo and Kwang-Sei Lee proposed by studying ferroelectricity and antiferromagnetism in multiferroic materials [13]. Recently, they carried out their studies by using the second quantization formalism for calculation of an impulsive polarization of ferroelectrics [14]. Due to nonlinear and complex media of the ferroelectrics, they display numerous types of nonlinear behaviors.. We investigate the possibility of complex chaotic vibrations propagating in **Barium Titanate Nanoparticles** using derivation of semiquantal dynamics via the Ehrenfest theorem.

**2- Extended Hamiltonian microscopic model of Barium Titanate**

Photon field couplings are formed between many-body electrons. In our study we will focus on photons whose finite mass is made up of interacting electrons. The photon field correlations of the electrons are the propagators of ferroelectric ordering due to frequent virtual absorption and emission of photons by electrons. It is revealed by many authors that multiferroics, spin glasses and polar glasses are assumed as composed of electron clusters. That is why they may be looked as finite block spins [15–20].

The effective Hamiltonian associated to **Barium Titanate** ferroelectric ordering can be characterized as an ensemble of effective photons for finite sized block spins in a tunneling potential of a double well assumption as follow

$$\hat{H} = \frac{\hat{P}^2}{2mN} + \frac{1}{2}k^2(\hat{x}^4 - 2x_0^2\hat{x}^2 + x_0^4), \quad k = Nm\omega^2 \qquad (1)$$

where $\hat{P}$ is the photon's momentum and $\hat{x}$ is the photon's position. The mass of the electron - photon is $m$, $N$ represents the average number of electro-photons in a finite standard block spin and $\omega$ represents the tunneling frequency for a double well potential appropriate to the KDP-type crystals. The repulsion between the core ion-ion of neighboring atoms leads to obtaining a periodic potential relative to the lattice site

Further, based on Ehrenfest's theorem [21], we will use the method of derivation of semi-quantal dynamics due to the fact that its approach is closer to the formulations of non-equilibrium static mechanics. Thus, the dynamics obtained is similar to the TDVP technique. Let us consider the motion of an effective photon for finite sized block spins in a one-dimensional time-independent tunneling potential. Its Hamiltonian will be $\hat{H} = \frac{\hat{P}^2}{2mN} + V(\hat{x})$ where by $\hat{O}$ are represented the operators. The centroid of a wave group representing the effective photon has the movement equations as follows:

$$\frac{d}{dt} < \hat{x} > = < \hat{p} >, \qquad \frac{d}{dt} < \hat{p} > = -< \frac{\partial V(\hat{x})}{\partial \hat{x}} >, \qquad (2)$$

where the <> indicate expectation values. In general, the centroid deviates from the classical trajectory. Using the following identity, we will expand the equations around the centroid:

$$< F(\hat{u}) > = \frac{1}{n!} < \hat{U}^n > F^{(n)}, \quad n \geq 0 \quad ou \quad F^{(n)} = \frac{\partial^n F}{\hat{U}^n}|_{<\hat{u}>} \; et \; \hat{U} = \hat{u} - < \hat{u} >. \quad (3)$$

Using the index summation convention and taking into account the commutativity property of the operators, we will be able to generate an infinite number of equations corresponding to an infinite Hilbert space for the considered problem. The assumption that the wave group is a squeezed coherent state renders the space finite will lead us to the following relations:

$$< \hat{X}^{2m} > = \frac{(2m)! \, \mu^m}{m! \, 2^m} \; (no \; summation), \quad < \hat{X}^{2m+1} > = 0,$$

$$4\mu < \hat{P}^2 > = \hbar^2 + \alpha^2 \quad and \quad < \hat{X}\hat{P} + \hat{X}\hat{P} > = \alpha, \qquad (4)$$

We can see that the above relations are similar to the generalized Gaussian wave functions [22-25,26]. (This assumption is precisely that of the TDVP: The wave group is confined to a given subspace). In addition, we introduce the change of variables $\mu = \rho^2 \; et \; \alpha = 2\rho\xi$.

However, the explicit form of the dynamics equation from the photonic tunneling model in ferroelectrics yields as follow

$$\frac{dx}{dt} = \frac{p}{mN}, \qquad 5a$$

$$\frac{dp}{dt} = 2k^2 x_0^2 x - 2k^2 x^3 - 6k^2 x \, \rho^2, \qquad 5b$$

$$\frac{d\rho}{dt} = \xi, \qquad 5c$$

$$\frac{d\xi}{dt} = \frac{1}{4\rho^3} + \rho(2k^2 x_0^2 - 6k^2 x^2) - 6k^2 \rho^3. \qquad 5d$$

The dynamics reduced of $x$ and $p$ are written for $<\hat{x}>$, $<\hat{p}>$ and are exactly those derived from the action principle [27]. The above equations are the Ehrenfest theorem ones we have set $\hbar = 1$. Due to the nonlinearity of the equations, we expect the system to have a chaotic behavior, the trajectories being both regular and irregular. it is worth noting that our equations are coupled, thus highlighting the connection between classical and quantum interactions. At $\hbar \to 0$ classical limit, only the first two equations remain, confirming that the fluctuation variables are responsible for quantum effects. Besides, the associated Hamiltonian is

$$H_{ext} = \frac{P^2}{2} + \frac{\xi^2}{2} + V_{ext}(x,\rho), \tag{6a}$$

$$V_{ext}(x,\rho) = \frac{1}{2}k^2(x^4 - 2x_0^2 x^2 + x_0^4) + \frac{1}{8\rho^2} + k^2\rho^2(3x^2 - x_0^2) + \frac{3}{4}k^2\rho^4, \tag{6b}$$

In the above equations the subscript *ext* is used in order to express the "extended" Hamiltonian and potential, respectively. This approach is very interesting, offering an explicit way of presenting the gradient system, and for a simplistic visualization of the semiquantal space, the extended potential is used. In this way, before the detailed numerical analysis, we can have an overview of the semiquantitative dynamics.

### 3. Numerical Simulations and discussions

In this section, the numerical analysis of equations (5) is addressed using the Runge-Kutta numerical integration method of the 4th order. The numerical interpretation was made on several sets of values for the parameter $k$ and different initial conditions. The results of the numerical analysis are obtained for the value $x_0 = 10$ when the quantum effects are small, but obviously they cannot be neglected.

For the different parameters, k, the periodicity of the system is determined, thus highlighting the shape of the effective photon trajectories. Figure 1 shows the irregular chaotic movement for the case where the tunnelling frequency admits low values.

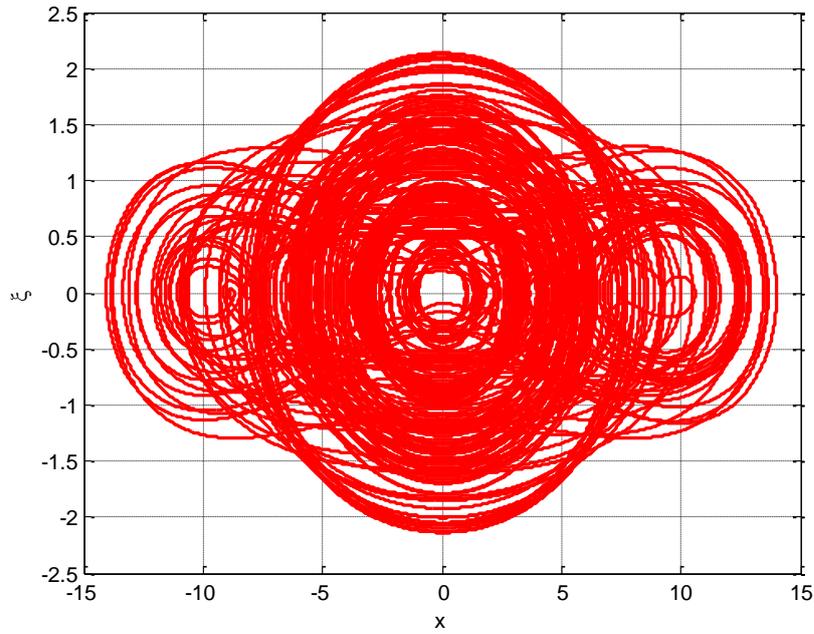

**Figure 1:** Poincare sections in the plane $(\xi, x)$. We set parameters as $k = 0.22$; $x_0 = 10$; $m = 0.00278$; with the initial conditions $(0.3; 0.3; 0.01; 0.01)$.

In order to determine the qualitative changes in the system dynamics when changing the values of parameter $k$ as well as the type of bifurcation and their nature, we performed simulations by varying parameter $k$ from 0 to 0.2. The Fig. 2 represents the evolution of the $x$-coordinate as a function of the parameter $k$.

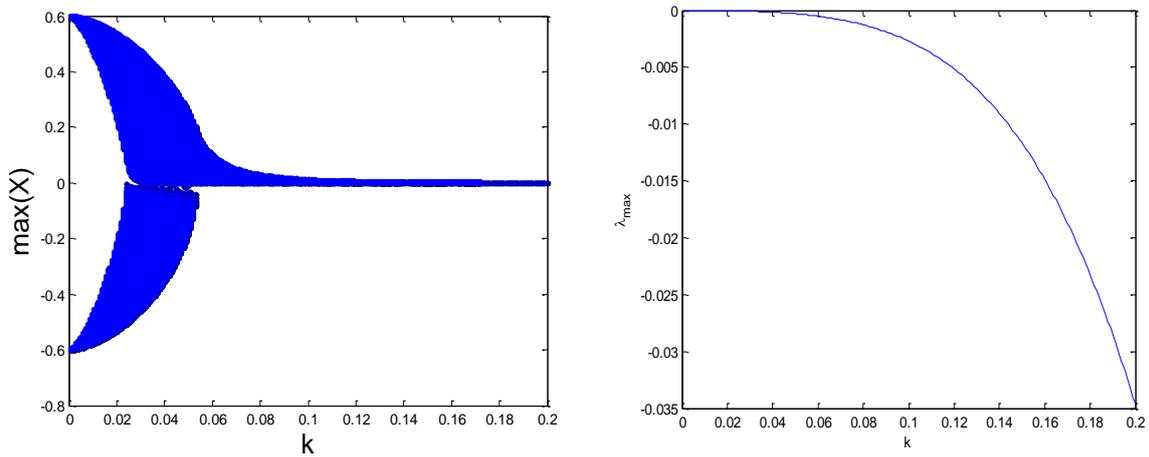

**Figure 2:** Dynamics with respect to the bifurcation diagram for $x_0 = 10$ under the effect of the tunneling frequency $\omega$ and the maximal Lyapunov exponent for the set of Eqs. (5) versus the parameter $k$ with the initial conditions $(0.3; 0.6; 0.1; 0.1)$.

During the evolution of the system as a function of parameter $k$, the behavior of the system undergoes two qualitative changes:

- For $k < 0.05$ the system contains no stable fixe point and then becomes chaotic. The state of the system changes completely. The trajectory always revolves around the attractor and the calculated coordinates are then arbitrary. A change in behavior occurs from $k = 0.05$. The equilibrium point is no longer a stable point, and the system stabilizes at one of the two points fixes.

- For $k > 0.05$ the trajectory converges towards the equilibrium point. The system is stable around this equilibrium point. Bifurcation is the event in which the qualitative property of attractor of a dynamical system is changed as a control parameter of the system is varied.

The calculation of the Lyapunov exponent is done from a numerical simulation and the logarithmic difference between the evolutions of the state variables of the system, the results obtained are represented by the Fig. 2. From Fig. 2 analysis, the Lyapunov exponent estimated is positive when the parameter $k$ is below 0.05, a positive exponent implies the divergence of neighboring trajectories, i.e. for these values of the system parameters, and we have a chaotic behavior. In addition, the Lyapunov exponent is negative when the parameter $k$ is over 0.05 that is to say the system is stable in this domain.

## 4. Conclusions

In this work, we derive an extended Hamiltonian photon correlation concept for multiferroics from conventional theories of pseudo-spin formalism and proton tunneling model. Because of the fact that ferroelectrics are nonlinear and complex media; their phenomena originate from spontaneous photon-phonon correlations to induce numerous kind of nonlinear behaviors. Order-disorder nature of phase transitions propagates in Barium Titanate ferroelectric through this study. In conclusion, chaotic regions and periodic regions exhibited by photon correlations are of course in certain correspondence with bifurcation diagram and Lyapunov exponent as shown in Fig. 2.

**References**

1. K. Werner, Solid State Physics. 4 (Academic Press, 1957).
2. M. Lines; A. Glass, (Clarendon Press, Oxford, 1979).

3. See J. Valasek, Physical Review. 15 537 (1920).

4. Safa Kasap, Peter Capper, (Springer Science+Business Media, 2006).

5. G. H. Haertling, Journal of the American Ceramic Society 82, 797 (1999).

6. Y. Reznikov, et al., Applied Physics Letters 82, 1917 (2003).

7. G. Ciofani, L. Ricotti, and V. Mattoli, Biomedical Microdevices 13, 255 (2011).

8. G. Ciofani, et al. in Piezoelectric Nanomaterials for Biomedical Applications 213 (Springer Berlin Heidelberg, 2012).

9. Aim Peliz Barranco, Advances in Ferroelectrics, InTech, (2012).

10. L. Lam, Introduction to nonlinear Physics, (Springer-velag, New York 1997).

11. F. C. Moon, Chaotic Vibrations - An Introduction for Applied Scientists and Engineers (John Wiley & Sons, New York, 1987); Chaotic and Fractal Dynamics -An Introduction for Applied Scientists and Engineers (John Wiley & Sons, New York, 1992).

12. Kazunori Aoki, Nonlinear Dynamics and Chaos in Semiconductors, (IOP Publishing Ltd, London, 2001).

13. J. H. Koo and K-S. Lee, Ferroelectrics Letters Section, 44, 42 (2017).

14. J. H. Koo and K-S. Lee, Ferroelectrics, 540, 4 (2019).

15. K. H. Fischer, and J. A. Hertz, Spin Glasses, (Cambridge University Press, New York, 1991).

16. K. Binder, and A. P. Young, Rev. Mod. Phys. 58, 801 (1986).

17. S. F. Edwards, and P. W. Anderson, J. Phys. F 5, 965 (1975).

18. D. Sherrington, and S. Kirkpatrick, Phys. Rev. Lett. 35, 1972 (1975).

19. Y. K. Kim, J. H. Koo, and K.-S. Lee, Ferroelectrics 494, 110 (2016).

20. K.-S. Lee, J. H. Koo, and C. E. Lee, Solid State Commun. 240, 10 (2016).

21. A. K. Pattanayak and W. C. Schieve, Physical Review Letters, 72, 2855 (1994); Physical Review Letters, 46, 1821, (1992). Proceedings from workshop in honor of Sundarshan., E.G.G. Gleeson, A.M., Ed., (world scientific, Singapore, in press).

22. F. Cooper, S.-Y. Pi and P.N. Stancioff, Phys. Rev. D 34, 3831 (1986).

23. A. Kovner and B. Rosenstein, Phys. Rev. D 39, 2332 (1989)

24. J. Klauder and B.-S. Skagerstam, 'Coherent States: Applications in Physics and Mathematical Physics' (World Scientific, 1985).

25. W.-M. Zhang, D.H. Feng and R. Gilmore, Rev. Mod. Phys. 62, 867 (1990).

26. Y. Tsue, Prog. Theor. Phys. 88, 911 (1992) and references therein.

27. E. Godwe, J. Mibaile, B. Gambo, S. Y. Dokac and T. C. Kofane, Chinese Journal of Physics 60, 379 (2019)